\documentclass{aastex}
\usepackage{emulateapj5}

\usepackage{natbib}
\newcommand{\Msolar}{{$\rm\ M_\odot$}}
\newcommand{\Msolarm}{{\rm M}_\odot}
\newcommand{\Msun}{\mbox{$M_{\odot}$}}

\newcommand{\de}{$^{\circ\ }$}

\begin{document}
 
\title{Using Pulsars to Detect Massive Black Hole Binaries
 via Gravitational Radiation: Sagittarius A$^*$ and Nearby
 Galaxies} 
 
\author{ A. N. Lommen and D. C. Backer}
\affil{Astronomy Department \& Radio Astronomy Laboratory\\
University of California, Berkeley, CA 94720-3411\\
email: alommen@astro.berkeley.edu, dbacker@astro.berkeley.edu}
 
\begin{abstract}
Pulsar timing measurements can be used to detect gravitational
radiation from massive black hole binaries.
The $\sim$106d quasi-periodic flux variations in Sagittarius A$^*$ 
(Sgr A$^*$) at radio wavelengths reported by \citet{Zhao01} 
may be due to binarity of the massive black hole that is presumed 
to be responsible for the radio emission. 
A 106d equal-mass binary black hole is unlikely
based on its short inspiral lifetime and other arguments. 
Nevertheless the reported quasi-periodicity has led us to consider 
whether the long-wavelength gravitational waves from a conjectured binary
might be detected in present or future precision timing of millisecond pulsars. 
While present timing cannot reach the level expected for an equal-mass
binary, we estimate that future efforts could. This inquiry has led
us to further consider the detection of binarity in the massive black
holes now being found in nearby galaxies. For orbital periods of
$\sim$2000d where the pulsar timing measurements are most precise,
we place upper limits on the mass ratio of binaries as small as
0.06. 
\end{abstract}

\section {Introduction}
\setcounter{footnote}{0}
\setcounter{figure}{0}

\citet{Sazhin78} and \citet{Detweiler79} discussed the influence
of long-wavelength (nanoHertz) gravitational radiation on the propagation 
of pulsar signals. \citet{Detweiler79} suggested a possible source
of such radiation: binary massive black holes (MBHs) in distant galaxies.
We have been engaged in a program to detect the stochastic background
from the Universe of coalescing MBHs as well as to
make estimates of the expected level. Here we consider the detection
of gravitational radiation from the nearest objects.

We begin this inquiry by considering our Galactic Center (GC).
There has been mounting evidence that
the dark mass detected via proper motions of IR stars
in the vicinity of Sagittarius A$^*$ (SgrA$^*$) is a MBH 
\citep{Eckart97, Ghez98, Maoz98, Ghez00}. Proper motion and absolute
astrometry techniques lead to the identification of the compact
non-thermal radio source Sgr A$^*$ with the MBH 
\citep{Backer99, Reid99, Menten97}. The electromagnetic emission of
Sgr A$^*$ may be either from the faint glow of matter
being accreted on the MBH or from cooling in a small disk/jet system 
(see e.g. Narayan, Igumenshchev \& Abramowicz 2000; Falcke 1996
\nocite{Narayan00, Falcke96}).
At mm and short cm wavelengths where the effects of interstellar
scattering are minimized the intrinsic source has been determined
to be of order a few AU, less than 100 Schwarzschild radii of
the MBH \citep{Lo98, Doeleman01}.

Recently, Zhao et al. (2000) reported quasi-periodic flux variations of 
Sgr A$^*$ with a 106d period using observations at 1.3 and 2.0 cm 
from the VLA. The authors explored various models to account for the 
variation including the possibility that the periodicity is related 
to the orbit of a binary companion. The authors discount
the binary scenario because the two
holes would be easily resolved by high angular resolution VLBI imaging assuming
that both were luminous. VLBI observations of Sgr A$^*$ at 22-43 GHz
reveal only a single source (e.g.,
Bower \& Backer 1998; Lo et al. 1998; Doeleman et al. 2001 
\nocite{Bower98, Lo98, Doeleman01}).
One could further argue that owing to the short lifetime for coalescence
of an equal-mass binary MBH in Sgr A$^*$ the system is unlikely.
The residuals in proper motion measurements of Sgr A$^*$ can also
be used to place an orientation dependent limit on the mass of
a dark companion to the Sgr A$^*$ MBH.
While binarity of the Sgr A$^*$ MBH is an unlikely explanation
for the flux variations reported by Zhao {\it et al.}, we proceed
in this paper to explore the detectability of the gravitational radiation 
from such a binary in millisecond pulsar (MSP) timing residuals.

The ratio of the hole masses now being measured in nearby galaxies
\citep{Magorrian98, Merritt01}
to their distance is such that these objects are also candidate sources
for detectable gravitational radiation if we make the binary hypothesis
for them also. In this case we have no candidate period and are free
to explore the limits on binary mass ratio at orbital periods where we are 
most sensitive, 2000 days. 

In this article we first discuss in \S 2 the possible amplitude of perturbation
of pulsar timing residuals by the conjectured Sgr A$^*$ binary MBH,
including a discussion of possible mass ratios.
In \S 3 we present our recent observations of MSPs and available archival
data. This is followed by a periodogram analysis that yields the
best limit we can reach with current data sets. 
\S \ref{sec:MDO}
discusses the possibility of detecting binary MBHs in nearby galaxies for
which hole masses have been estimated. 
In our final section we summarize our conclusions.

\section {Perturbation of Pulsar Timing by Gravitational Radiation from Sgr A*} \label{sec:perturbation}

We use the superb formulation of the gravitational radiation
from binary masses by \nocite{Peters63} Peters \& Matthews (1963, 
hereafter PM63) to make detailed estimates of the amplitude
$h(\vec r,t)$ of the possible gravitational emission from Sgr A$^*$.
We then follow the development by \citet{Sazhin78} and \citet{Detweiler79} of
the influence of this radiation on the electromagnetic pulses
emitted from a pulsar as it travels through space-time that
is perturbed by $h(\vec r,t)$. In short, pulse propagation 
through complete cycles of $h(\vec r,t)$ have no net effect on 
the arrival time. There
is only a perturbation of the arrival time by the
incomplete cycles traversed at the pulsar and at Earth.

We use Eqn. 16 in PM63 to calculate the luminosity $L_{GW}$
of the gravitational wave (GW) from the assumed circular binary system
in the GC

\begin{equation}
  \label{eqn:L1}
  L_{\rm GW} = {32\over 5}{m^5q^2\over a^5(1+q)^4}{G^4\over c^5}
     =  2.9\times 10^{43}{\rm~erg~s}^{-1}{q^2\over(1+q)^4}
\end{equation} 
where $m=2.6\times 10^6$ \Msun~ is the total mass of the 
system\cite[]{Ghez00}, $q$ is the mass ratio ($q\le 1$), 
and $a$ is the semi-major axis. 
The numerical result uses 
Kepler's Law $a^3=GmP_{orb}^2/4\pi^2$ and 
the 106d orbital period of interest. For $P_{orb}=106d$, $a=59{\rm~AU}$.
Here and below we assume a circular
orbit which is likely following the combined actions
of dynamical friction and radiation.

From the energy density of a GW, which is 
$U = c^2\dot{h}^2/32\pi G$ (Eqn. 2 in PM63), we derive 
the dimensionless amplitude of the GW;
\begin{equation}
h  = {34} {\left({m^{1.67}G^{1.67}}\over{P_{orb}^{0.67}d}\right)} {q\over(1+q)^2} = 6.3 \times 10^{-14} {q\over(1+q)^2},
\end{equation} 
where $d$ is the distance to the emitter, 8 kpc. In this expression
$h$ is averaged over all orientations of the observer relative
to the plane of the binary orbit.  
Since we obtained this
expression from the total energy density, which is the sum of the
contributions from two polarizations, 
$h_+$ and $h_\times$, 
$h$ is actually the quadrature sum of
these two polarizations.
In order to find the dependence
on inclination angle, $i$, we use the
expression for the average power radiated per solid angle in PM63.
This shows that the power radiated along the
axis of the orbit is 8 times that for an edge-on view.

As discussed by \cite{Detweiler79} and others the dimensionless strain $h$
produces an apparent redshift in the pulsar frequency. A periodic source
of GW then will produce a periodic
shift in pulse arrival time from propagation through the
gravitational radiation with an amplitude, $\delta t$, which is given by
\begin{equation}
\label{eqn:angular}
\delta t  \sim {hP_{\rm GW}\over{2\pi}}= 22 {\rm~ns}\sqrt{(1 
+ 6\cos^2i + \cos^4i)}{q\over (1+q)^2},
\end{equation}
where $P_{\rm GW}$ represents the period of the gravitational
wave, $P_{orb}/2$.
The mass ratio factor is at most $1/4$.  
The angular factor ranges from 1 to 2.8.
Therefore, $\delta t$ is less than 16 ns, and its average value over all
solid angles is
\begin{equation}
\label{eqn:46ns}
\delta t  \le  46 {\rm~ns}{q\over (1+q)^2},
\end{equation}
which is 11 ns for $q=1$ and less for any other mass ratio.

\cite{Detweiler79} discusses the dependence of
the GW signature in pulsar timing on the angle between
the GW and the pulsar sightline. 
Pulse propagation times are perturbed by the GW
owing to incomplete traversal of a cycle of the GW 
both at the pulsar as pulses are emitted and
at the Earth upon pulse reception.
In the plane wave approximation the resulting timing residual, $\delta t$,
is
% This equation follows from Detweiler (29) and (10)
\begin{equation}
\label{eqn:detweiler}
  <{{\delta t}\over P_{\rm GW}}>={1\over{2\pi}}\left[{(1+\gamma)\over 2}
  h\right]
  (f(ct_{\rm r}) - f(ct_{\rm e}+\gamma l)).
\end{equation}
where $\gamma$ 
is the cosine of the angle, $\phi$, between the GC and the pulsar, where 
$\phi=0$ is defined as the pulsar lying along the line of sight to the GC.
$f(t)$ is the dimensionless phase term that comes from 
the fragments of the GW traversed
at the emitter (e) and receiver (r) ends ($f\le1$). 
The times of 
emission and reception are $t_{\rm e}$ and  $t_{\rm r}=t_{\rm e}+l/c$,
respectively, and the factor $\gamma l$ is the projection
of the pulsar distance $l$ along the GW
propagation vector.  Note that 
when $\gamma=+1$, there is no effect from the GW: a pulsar
lying along the line of sight to the GC
will experience no effect.
The residual is also identically zero for $\gamma=-1$
which describes electromagnetic
waves (EMW) traveling
in the opposite direction as the GW.
The residual increases as the EMW and GW become perpendicular, and
reaches a maximum just before they become parallel.
The angle between PSR B1937+21 and Sgr A$^*$ from Earth 
is 58\de, and between
PSR J1713+0747 and Sgr A$^*$ is 37\de.  
The signal from the two pulsars will therefore be diminished 
from our earlier estimate in Equation \ref{eqn:46ns} by a factor
$(1+\gamma)/2 = 0.76$ and 0.90, respectively.
This factor adjusts our earlier estimate of the maximum possible
effect from 16 ns to 14 ns, 
and our estimate of the average effect
from 11 ns to 10 ns due to an equal-mass binary.

If either of the sources were much closer
to the GC than the Earth, then
$h$ in Eqn. \ref{eqn:detweiler} would need
to be corrected for both the different wave amplitudes
and the different angular factors.
In our case
all relevant distances (Earth to GC, J1713+0747
to GC, B1937+21 to GC) are 7-8 kpc, and are not known
to better than 25\%~\citep{Kaspi94, Camilo94}.  Since the
distances and therefore the phase factors are unknown, and
the amplitudes of the two effects at the two ends are roughly
equal, it is even possible that the two phase factors
will be nearly the same, vastly diminishing
the signal.
The signal at emitter
site will represent GW level $10^4$ years ago, probably
not too different from today.
The phase factor from the receiver end
will produce a signature that is correlated with other pulsars,
whereas the emission terms will be uncorrelated.
Clearly observations with an array of pulsars at different angles
and distances are critical to overcoming this ``emission phase" noise.

We have used a mass ratio
of $q=1$ to calculate the maximum signal we might expect.
There are two arguments against a large mass ratio.  One is based on
evolutionary arguments and the other on proper motion observations
of Sgr A$^*$.  First, the lifetime for
gravitational inspiral from a 106d orbit for $q=1$ is $3\times 10^6$ y;
the lifetime increases with $(1+q)^2/q,~q\le 1$.  
After two galaxies merge the timescale for their central black holes 
to reach such an orbit is unknown, and could be from 30 million years to
a Hubble time (e.g., Rajagopal \& Romani 1995, Gould \& Rix 2000 
\nocite{Gould00, Rajagopal95}).  \cite{Toth92} rule out the possibility
of a recent merger using
models of disk heating via accretion of satellite galaxies. They demonstrate
that no more than 4\% of the mass of the galaxy could have been accreted
within the last 5 billion years.  \cite{Xu94} model the
accumulation of a central MBH in a scenario with much less disk heating:
the accretion of primordial $\sim 10^6~\Msolarm$ black holes.
They show that a quickly accumulating MBH in the GC is actually not what we 
should expect;  dynamical friction and gravitational radiation serve to
eject massive objects from the center as well as accreting them.  Galaxies
such as ours, they conclude, are usually host to zero, one or two MBHs.
In any case, a crude estimate of the likelihood that we happen to
be living in the epoch of coalescence of a $2.5 \times10^6 \Msolarm$ 
106d binary is $\sim3\times10^6$y/(age of Milky Way) or less than 0.1\%;
in other words it is unlikely that we are observing the pair of
black holes at the moment of their coalescence.

The second argument uses the VLA and VLBA measurements
of the proper motion of Sgr A$^*$
\citep{Backer99, Reid99} to place a limit on the mass ratio. 
The three reported VLBA measurements span almost
exactly 7 cycles of the observed 106d quasi-periodicity. 
The
middle point is offset in phase from the end points by 0.5.  The scatter
about the line connecting the VLBA measurements, about 0.5 mas,  
provides an estimate of the maximum separation between the observed 
mass and the center of mass of the system along the axis where
the observational data are most sensitive.  If we assume Sgr A$^*$
is centered on the larger of the two masses, this places a limit on
the mass ratio of $q < \rm(0.5{\rm~mas)}({8\rm~kpc})/59{\rm~AU} = 0.07$.
This limit is mainly in the EW direction where the VLBA data are most 
precise. Thus, the limit is strictly on $q \sin(\theta)$ where $\theta$ 
is the inclination of the orbital plane to the plane defined by the NS
direction and the line of sight.
Interestingly, the VLBA data nearly rule out the possibility that we are 
seeing the less massive member of the conjectured binary.

In conclusion, we estimate that the perturbation in pulsar timing
residuals is no larger than $\sim$ 14 ns for the pulsars we consider here.
In the following section, we present the current level of precision in
our measurements, and discuss the feasibility of detecting a 14-ns
sinusoidal amplitude in our timing residuals.

\section {Observations} \label{sec:observations}

We have been conducting monthly observations at 0.43 GHz,
1.4 GHz and 2.4 GHz of an array of
MSPs using the Arecibo Observatory 300 m 
telescope\footnote{The National Astronomy and Ionosphere Center
Arecibo Observatory is operated by Cornell University
under contract with the National Science Foundation.}
since December 1997. These data are used
to make precision arrival time measurements for a variety
of astrophysical goals. We used the Arecibo-Berkeley
Pulsar Processor (ABPP), which is a multi-channel, coherent dispersion 
removal
processor\footnote{`coherent' means that the dispersion is removed in
the voltage domain prior to power detection.} 
with 112-MHz total bandwidth capability.  

In this paper we include only PSRs J1713+0747, B1855+09 and B1937+21 
because their data sets span the longest time, 3.2 years, 
and they have the best timing precision.
For PSRs J1713+0747 and B1855+09 
we use 56-MHz bandwidth for observations at 1.4 GHz, and 112-MHz for 
2.4 GHz, while for PSR B1937+21 we use 45-MHz overall bandwidth at 1.4 
GHz and 55-MHz at 2.4 GHz.
Calibrated total intensity profiles were formed from signals with
orthogonal circular polarization.
The profiles were then cross correlated with a template to
measure times of arrival (TOAs) relative to the observatory atomic clock.
Small errors in the observatory UTC clock,
of order 1 $\mu$s, were corrected based on
comparison of local time to transmissions from the Global
Positioning System of satellites (GPS) using the Totally Accurate
Clock receiver at the observatory.
GPS time is then corrected to the TAI scale via
publications from the Bureau International des Poids et Mesures (BIPM).
The resulting TOAs are modeled for spin, astrometric, and, when relevant, 
binary parameters using the TEMPO software
package\footnote{http://pulsar.princeton.edu/tempo}. In this paper
we use the residuals from the model to look for other effects.

\begin{figure}
\plotone{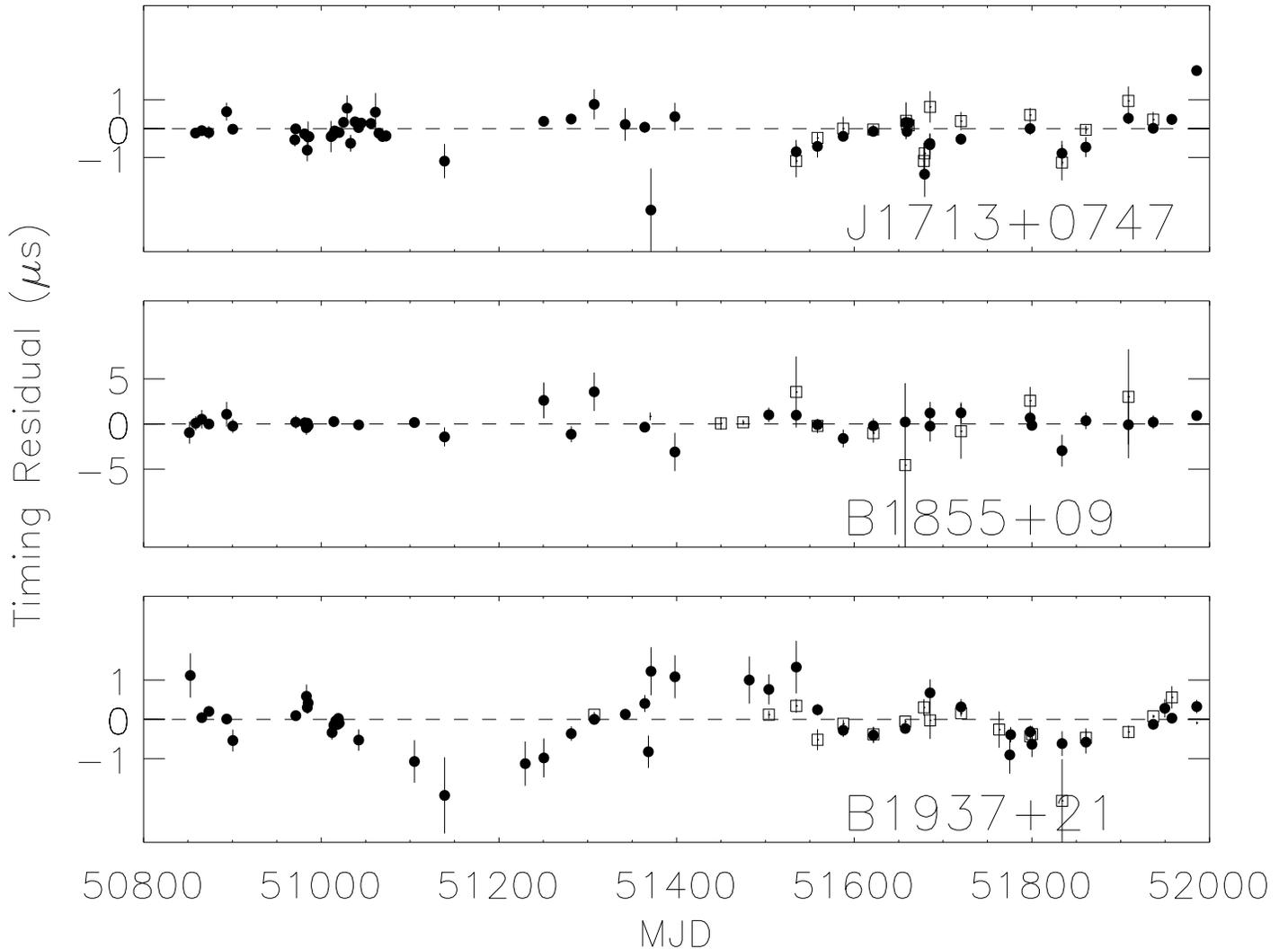}
\caption[]{
\label{fig:TOAS}
Timing residuals for PSRs J1713+0747, B1855+09 and B1937+21
using our data taken after the Arecibo upgrade.
Filled circles are 1420-MHz data and squares
are 2380-MHz data.
}
\end{figure}

After fitting for phase, spin period ($P$), period derivative ($\dot{P}$), 
right ascension, declination, and proper motion in the
B1937+21 data, and
additionally for the 5 Keplerian binary parameters in PSR J1713+0747 and
PSR 1855+09, we have the residuals shown in Figure \ref{fig:TOAS}.
Note that we did not fit for the Shapiro delay which has been
measured in both PSRs J1713+0747 and B1855+09, 
but rather set the values to the best-fit values published 
\citep{Kaspi94, Camiloshapiro}.
The weighted RMS of the
day-averaged residuals are 
0.35 $\mu$s,
0.53 $\mu$s, and
0.14 $\mu$s for J1713+0747, B1855+09, and B1937+21, respectively.

In our periodogram analysis (\S \ref{sec:periodogram}) 
we also use the Princeton Mark II 
and Mark III data published and
made public by Kaspi, Taylor, \& Ryba (1994, hereafter KTR94)
\nocite{Kaspi94}.  This was a biweekly monitoring
program that spanned 8 years.  KTR94 carefully removed dispersive effects
from each of their TOAs on PSR B1937+21, which we have not
done in the ABPP data.  Due to our large bandwidth (45 MHz vs their 10 MHz) we
achieve a similar level of precision with shorter data span.  

PSR B1937+21 has been demonstrated by KTR94 to be unstable
on time scales of about 5 years, but on the time scales over which we
are interested here, 25-100 d, any instability it may have is below
the noise level of our data, about 0.14 $\mu$s.

The timing measurements at Arecibo Observatory
have a short-term precision of order 50 ns on
the ``best''  millisecond pulsars with integrations of 3 minutes
over a single hour. This suggests that 
with sufficient averaging and frequent observation we 
could detect the 14-ns amplitude 
perturbation discussed above.
However, we have not achieved this level of precision over the full
data span of our current data ( $\sim$ 3 y).
We suspect that the discrepancy is the result of distortion of
the average pulse profile by diffractive
scintillation across our wide band (50-110 MHz).
The pulse profile
evolves with frequency over the band, and diffractive scintillation
weights different parts of the band more heavily on different
days.  We are working on an algorithm to suppress this effect
\citep{Lommenpevol01}.

Nevertheless we 
proceed in \S \ref{sec:periodogram}
with a careful analysis of the current limits our data 
can place on the existence of a GC binary MBH.

\section {Periodogram Analysis} \label{sec:periodogram}

The residual data shown in Figure \ref{fig:TOAS}
represent an irregularly spaced, sparsely sampled system
in which we look for periodicities.  As such, the method described
by \nocite{Cumming99} Cumming et al. 1999 (hereafter C99) for constructing 
and normalizing periodograms is ideal.  
C99 were searching for planets in spectroscopic
velocity data from the Lick Observatory.  The astronomical goal is very 
different, but the method is essentially identical.
This periodogram approach is similar to the approach taken by 
\cite{Bailes93} to look for planets around slow period pulsars.

We constructed periodograms for the residual timing data in the following
way.  We fit the data to the
function $A\cos{\omega t} + B\sin{\omega t} + C$ by minimizing $\chi^2$,
at a range of frequencies, $\omega=2\pi f$, centered at the nominal 
GW frequency, $f_0$ of 
1/53 d$^{-1}$.  The arbitrary
offset `C' is critical for sparsely sampled data as demonstrated by C99,
even though the mean of the residuals is fit by the period polynomial.
The range of frequencies we considered was dictated by the 
width of the periodogram feature detected by \citet{Zhao01} which 
was approximately
unity for the 2-cm data.  We chose to look at a frequency range, 
$\Delta f = \frac{1}{2}f_0$, which is 0.014 to 0.024 d$^{-1}$
corresponding to a range of periods from 42 to 71 days.
The frequency range considered in the analysis is very important
as the smaller the range considered for detection, the more
sensitive the measurement will be. 
In order to sample the possible periodicities completely, 
the approximate size of the steps by which we needed to sample
this frequency range is $1\over{S}$, where $S$ is the time span
of the data set.  Since we are using data sets of different spans
(namely the Kaspi data which is 8 years long, and the post-upgrade
data which is 3 years long)
we chose a frequency spacing that would slightly oversample the
Kaspi data, and 4 times oversample the post-upgrade data:
stepsize $=2\times10^{-4}$ d$^{-1}$.

\begin{figure}
\plotone{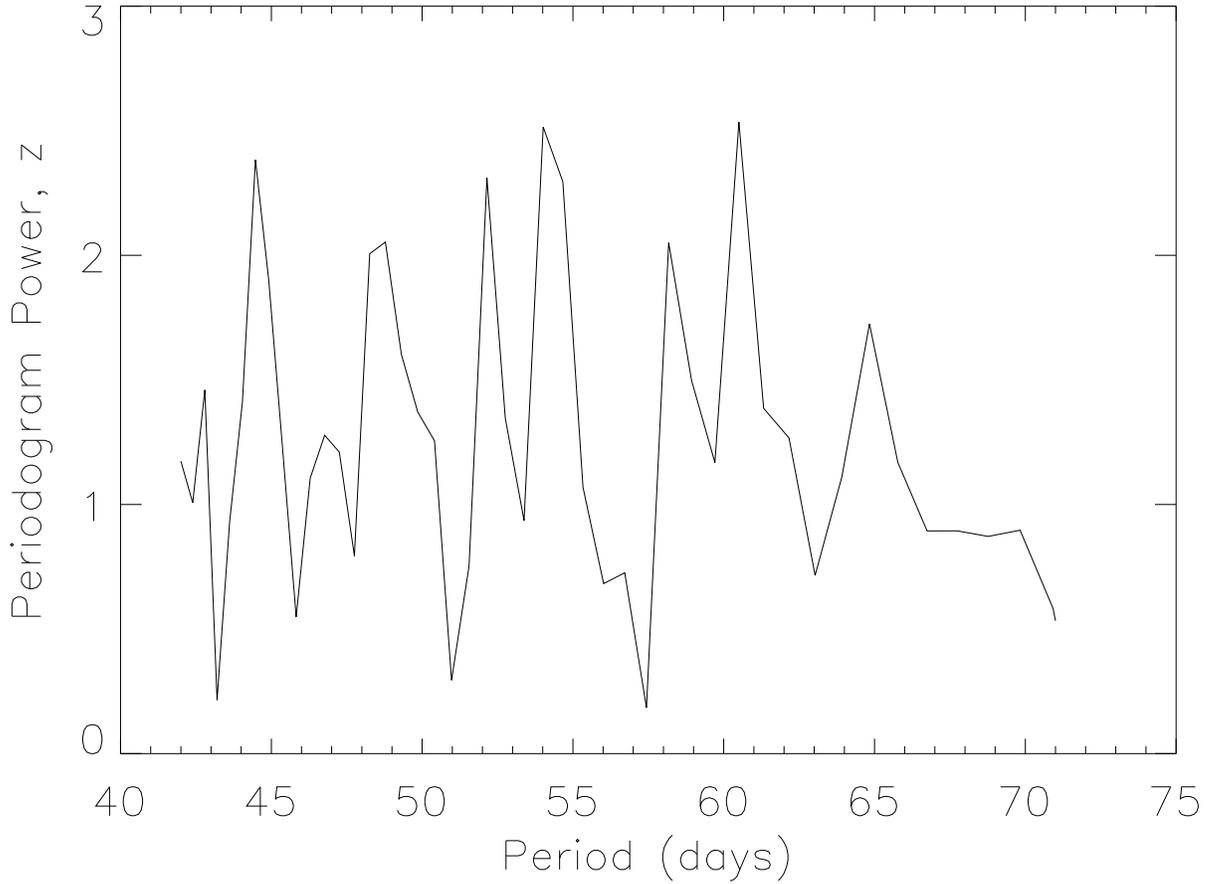}
\caption[]{
\label{fig:spectra}
Periodogram power, $z$, vs. period
in days as measured in residuals from PSRs B1937+21 and J1713+0747.
}
\end{figure}

In Figure \ref{fig:spectra} we use
the following formalism, taken from C99, to plot the periodogram
power $z(\omega)$ as a function of $\omega$.
\begin{equation}
z(\omega)=\frac{(N-m-3)}{2}\frac{(\chi^2 - \chi^2(\omega))}{\chi^2(\omega_0)}
\label{eq:z}
\end{equation}
where $N$ is the number of degrees of freedom of the data, $m$
is the number of independent fit parameters, $\chi^2$ is the
weighted sum of the squares of the original residuals, 
$\chi^2(\omega)$ is the
weighted sum of the squares of residuals after 
the periodicity fit is included and $\omega_0$ is the
frequency that gives the lowest $\chi^2(\omega)$.  
$N-m-3$ is the number of degrees of freedom of the periodicity fit and
corresponds to a residual fit using the $m$ parameters in the pulsar
model, and 3 additional parameters corresponding to fitting $A$, $B$, and $C$.
Therefore $z(\omega)$ is the
amount by which the $\chi^2$ is reduced by adding the periodic signal
to the data, normalized by the best fit $\chi^2$.
Note that $z(\omega)$ is normalized such that $z(\omega)=1$ for
no sinusoidal signal present in the residuals.

We decided not to include the data from PSR B1855+09 in our analysis
due to the higher RMS of its residuals compared to the other pulsars.
The normalization of
the statistic $z(\omega)$ allows us to average
the three periodograms from the 3 remaining data sets to acquire the
final periodogram shown in Figure \ref{fig:spectra}.
We also repeated the analysis using only PSR B1937+21 data
and achieved very similar results.

Is there any significant detection of gravitational radiation
in the periodogram shown in Figure \ref{fig:spectra}?
To answer this question we use Monte Carlo simulations
to determine whether the peak at 60.5 d, with a value of 2.5,  
is a spurious effect of the noise.  Following
the method described in C99, we created 400 sets of simulated residuals,
with no periodic signal, but with identical statistics to the real data.
We asked what percentage of these 400 realizations would conspire to produce
a peak between 42 and 71 days, as high or higher than 2.5.
This percentage is called the ``false-alarm probability", 
and if it is higher than some threshold the detection is spurious.  
The threshold must be determined by the specifics of the problem, and
can be anywhere from $10^{-5}$ to 0.1.
C99, for example, used a threshold of $10^{-5}$ because there were
systematics in their data that would imitate a signal at higher values.
Creating identical statistics to the original data set proved to
be challenging and also crucial to producing a meaningful
false-alarm probability.   In all cases we replicated the sampling
of the original data.  When we merely generated random numbers with
a gaussian noise distribution and the same RMS as the original data,
the false-alarm probability was 100\%.  This is due to the ``redness''
of the variation in the residuals, i.e., neighboring residuals have
much smaller RMS with respect to each other than the overall RMS of
the data set.  Therefore by creating gaussian noise with a particular RMS
we were creating a data set with much higher RMS between neighboring
points than was present in the original set.  The redness of
pulsar timing residuals is commonly called ``timing noise" and has
been studied and discussed by a number of authors, and will not
be discussed here~\citep{Cordes80, Arzoumanian94, Kaspi94}.
To account for this low-frequency variation we fit for 3 additional
period derivatives in each data set.  This is identical to removing
a 5th order polynomial from each data set.  We are confident that
this additional fitting would not remove any signal at a 53 day period
since a 5th order polynomial has 5 zero-crossings, and our data sets
are all at least 3 years long.  We may, therefore, be removing variations
as short as (3 yr/5) $\sim$ 200 days but no shorter.  
Additionally, to fabricate the data we merely randomly reordered the
real data, i.e. randomly assigned real residuals to the wrong dates,
to assure that the statistics were identical.  The final 
false alarm probability for producing the peak shown in Figure \ref{fig:spectra}
was 23\%, and thus the detection is spurious.
To be sure we are not washing out
features in the power-spectrum present in any one data set
we have performed this analysis
on each data set individually and obtained very similar results.

In order to find the minimum amplitude detectable in our data
we used Monte Carlo simulation in the following fashion.
We simulated timing residuals with the
same statistical properties as our real data using the same method
described above, and 
injected a signal
at a 53-day period at a variety of amplitudes.  The amplitude that
we designate as our `limit' corresponds to the amplitude where 99\%
of the random realizations of simulated data produced a power amplitude
larger than that which we actually measure.
This amplitude, $R = 0.15~\mu$s, represents the minimum amplitude we
can detect with our current data.

\section{Nearby Massive Dark Objects} \label{sec:MDO}
  
What is the probability
that massive dark objects (MDO) 
in other galaxies are effecting our timing residuals?
\cite{Magorrian98} determined the masses or mass limits of 29 MDOs in 
nearby galaxies
using HST photometry and ground-based spectroscopy of the galaxies.
Several independent studies have demonstrated that these estimates are 
systematically high\citep{vandermarel99, Wandel99, Ferrarese00}.  
\cite{vandermarel99} suggests the discrepancy
is due to the assumption of velocity isotropy made by Magorrian et al.
\cite{Merritt01} improved the mass determination by using the
extremely tight relationship between MDO and the velocity dispersion
of the host bulge.  In our analysis we use the Magorrian et al. 
sample of galaxies with the updated masses by \cite{Merritt01}.

We choose an advantageous orbital period for this discussion, 
2000 days, where our data place the most stringent limit, and the
lifetimes of the orbital systems are longer.
The wavelength of the GW emitted from a 2000-day system is 1000 days, which
is the length of our data set,
so we must be careful that our pulsar model fitting would not remove
the signal we wish to detect.  A third period derivative fit or 
higher order would do so.  Consequently, in order to obtain a meaningful
limit we did not fit for the additional 3 derivatives of the period as we
did for the previously described 
periodogram analysis, but rather used the residuals
shown in Figure \ref{fig:TOAS} which include only a single period derivative.
By the same technique described in \S\ref{sec:periodogram}
we determined that
the detectable amplitude at this period is 170 ns.
In Table \ref{tab:magorrian} we show the amplitude of the effect
on the timing residual from each of the objects studied by 
\cite{Merritt01} assuming the MDOs are equal-mass binaries with 2000-day 
orbits.  Only those which produce a residual amplitude greater
than 10 ns are tabulated.  The amplitude shown is averaged over all
solid angles.  The amplitudes are as large
as 850 ns which would be detectable in our data.  
However, the probability of detecting such an object during coalescence is
proportional to its lifetime.  
Lifetime, $\tau$,
scales with $q$, $M$, and $P_{orb}$ as follows.
\begin{equation}
  \label{eqn:lifetime}
  \tau = 8.0\times10^4{\rm~y}{(1+q)^2\over{q}}\left({m\over{10^9\Msolarm}}\right)^{-5/3}\left(P_{orb}\over{2000 d}\right)^{8/3}
\end{equation}
With these larger mass systems, we gain amplitude of the gravitational
wave ($\propto m^{5/3}$), but lose reasonable lifetime of the system
by the same factor.
Fortunately, lifetime also scales strongly with orbital period
in our favor ($\propto P_{orb}^{8/3}$), 
so we have much to gain by looking at longer periods.
We place upper limits on q for 
these objects in Table \ref{tab:magorrian} 
just as we have for the GC
using the 170-ns detectable amplitude.
These limits are as small as $q \le 0.06 $, which correspondingly
give much longer lifetimes for the systems as shown in the last column.
Objects with a reasonable lifetime ($\sim$1~Gyr) have a 
$P_{orb}$ too large to be detectable by the pulsar measurements.

\renewcommand{\arraystretch}{.7}
\begin{table}
\caption[]{
\label{tab:magorrian}
Pulsar Timing Residuals of Binary Massive Black Holes
}
\begin{small}
\begin{center}
\begin{tabular}{lrrrrrr}
\tableline
Object &   Mass\tablenotemark{a} & Distance & Residual\tablenotemark{b} & Lifetime\tablenotemark{b} & Limit to & Lifetime\tablenotemark{c}\cr
& $\left[{\log{m\over\Msolarm}}\right]$ &  [Mpc]  &  [ns] & [y] & Mass Ratio, q & [y] \cr
\tableline
  NGC1399 &  9.0 &  17.9 &      327 &   2.9E+05 &    0.18  &   5.7E+05 \cr 
  NGC1600 &  9.0 &  50.2 &      102 &   3.3E+05 &  \nodata &   \nodata \cr 
  NGC2300 &  8.7 &  31.8 &       57 &   9.3E+05 &  \nodata &   \nodata \cr 
  NGC2832 &  9.3 &  90.2 &      157 &   1.2E+05 &    0.66  &   1.3E+05 \cr 
  NGC3115 &  8.8 &   8.4 &      281 &   7.2E+05 &    0.22  &   1.2E+06 \cr 
  NGC3379 &  8.1 &   9.9 &       18 &   9.2E+06 &  \nodata &   \nodata \cr 
  NGC3608 &  8.2 &  20.3 &       10 &   7.6E+06 &  \nodata &   \nodata \cr 
  NGC4278 &  8.7 &  17.5 &      107 &   9.1E+05 &  \nodata &   \nodata \cr 
  NGC4291 &  8.7 &  28.6 &       63 &   9.3E+05 &  \nodata &   \nodata \cr 
  NGC4472 &  8.8 &  15.3 &      134 &   8.3E+05 &  \nodata &   \nodata \cr 
  NGC4486 &  9.2 &  15.3 &      845 &   1.3E+05 &    0.06  &   6.6E+05 \cr 
  NGC4552 &  8.7 &  15.3 &      119 &   9.3E+05 &  \nodata &   \nodata \cr 
  NGC4594 &  8.6 &   9.2 &      104 &   1.8E+06 &  \nodata &   \nodata \cr 
  NGC4621 &  8.3 &  15.3 &       26 &   4.2E+06 &  \nodata &   \nodata \cr 
  NGC4649 &  9.1 &  15.3 &      610 &   1.8E+05 &    0.08  &   6.6E+05 \cr 
  NGC4660 &  8.2 &  15.3 &       17 &   6.3E+06 &  \nodata &   \nodata \cr 
  NGC4889 &  9.4 &  93.3 &      256 &   7.1E+04 &    0.25  &   1.1E+05 \cr 
  NGC6166 &  9.0 & 112.5 &       50 &   3.0E+05 &  \nodata &   \nodata \cr 
\tableline
\end{tabular}
\end{center}
\end{small}
\tablenotetext{a}{From \cite{Merritt01}.}
\tablenotetext{b}{Assumes equal-mass binary and orbital period of 2000 d.}
\tablenotetext{c}{Assumes mass ratio shown in column 6 and orbital period of 2000 d.}
\end{table}

\section{Conclusion}

Gravitational radiation of
an equal-mass 2.5 $\times 10^6$ \Msolar~black hole
binary at the GC would produce a periodicity in
pulsar arrival times of order 10 ns.  While this is an order
of magnitude below the limits of present data presented
in this paper, in the future a special observing effort might 
reach such a detection level. However, published proper motion
measurements of Sgr A$^*$ place a limit on the 
mass ratio of any such binary of 20:1 for low inclinations and
EW orientation.
In this case the gravitational radiation effects would be
below detection limit in pulsar timing. A small mass ratio
is also likely given consideration of the coalescence time
scale and the absence of any evidence of a recent large mass accretion
event in the GC. 

The known MDOs in nearby galaxies, if binary MBHs
with orbital periods around 2000d, would produce
a larger signal, up to $\sim$ 1~$\mu$s, than that estimated 
for Sgr A$^*$. However
the lifetimes to gravitational radiation inspiral
for such binaries are shorter than the already short lifetime
of Sgr A$^*$ and therefore lower the probability that we
are seeing them in this phase of evolution.  With a
number of such objects in existence, the probability increases
that at least one of them is still in a `young' binary state,
and might be seen in pulsar timing residuals.
Maintaining of precision pulsar monitoring programs with long and continuous
coverage is important for the future of such detection efforts.
The ``Pulsar Timing Array", 
our precision millisecond pulsar timing program,
extends the work we have described here to include the entire
ensemble of MBH-MBH systems in the universe.  This ensemble
produces a stochastic GW background at a level that we can detect.
This measurement will place important constraints on
the origin and evolution of MBHs.

\acknowledgements
We are grateful to Andrew Cumming and Yuri Levin for very useful
discussions.
We thank the Arecibo Observatory telescope operators
for many days and nights of assistance,
and Dunc Lorimer for pioneering remote observing at AO.

\renewcommand{\arraystretch}{.7}
\begin{table}
\caption[]{
\label{tab:magorrian}
Pulsar Timing Residuals of Binary Massive Black Holes
}
\begin{small}
\begin{center}
\begin{tabular}{lrrrrrr}
\tableline
Object &   Mass\tablenotemark{a} & Distance & Residual\tablenotemark{b} & Lifetime
\tablenotemark{b} & Limit to & Lifetime\tablenotemark{c}\cr
& $\left[{\log{m\over\Msolarm}}\right]$ &  [Mpc]  &  [ns] & [y] & Mass Ratio, q &
[y] \cr
\tableline
  NGC1399 &  9.0 &  17.9 &      327 &   2.9E+05 &    0.18  &   5.7E+05 \cr
  NGC1600 &  9.0 &  50.2 &      102 &   3.3E+05 &  \nodata &   \nodata \cr
  NGC2300 &  8.7 &  31.8 &       57 &   9.3E+05 &  \nodata &   \nodata \cr
  NGC2832 &  9.3 &  90.2 &      157 &   1.2E+05 &    0.66  &   1.3E+05 \cr
  NGC3115 &  8.8 &   8.4 &      281 &   7.2E+05 &    0.22  &   1.2E+06 \cr
  NGC3379 &  8.1 &   9.9 &       18 &   9.2E+06 &  \nodata &   \nodata \cr
  NGC3608 &  8.2 &  20.3 &       10 &   7.6E+06 &  \nodata &   \nodata \cr
  NGC4278 &  8.7 &  17.5 &      107 &   9.1E+05 &  \nodata &   \nodata \cr
  NGC4291 &  8.7 &  28.6 &       63 &   9.3E+05 &  \nodata &   \nodata \cr
  NGC4472 &  8.8 &  15.3 &      134 &   8.3E+05 &  \nodata &   \nodata \cr
  NGC4486 &  9.2 &  15.3 &      845 &   1.3E+05 &    0.06  &   6.6E+05 \cr
  NGC4552 &  8.7 &  15.3 &      119 &   9.3E+05 &  \nodata &   \nodata \cr
  NGC4594 &  8.6 &   9.2 &      104 &   1.8E+06 &  \nodata &   \nodata \cr
  NGC4621 &  8.3 &  15.3 &       26 &   4.2E+06 &  \nodata &   \nodata \cr
  NGC4649 &  9.1 &  15.3 &      610 &   1.8E+05 &    0.08  &   6.6E+05 \cr
  NGC4660 &  8.2 &  15.3 &       17 &   6.3E+06 &  \nodata &   \nodata \cr
  NGC4889 &  9.4 &  93.3 &      256 &   7.1E+04 &    0.25  &   1.1E+05 \cr
  NGC6166 &  9.0 & 112.5 &       50 &   3.0E+05 &  \nodata &   \nodata \cr
\tableline
\end{tabular}
\end{center}
\end{small}
\tablenotetext{a}{From \cite{Merritt01}.}
\tablenotetext{b}{Assumes equal-mass binary and orbital period of 2000 d.}
\tablenotetext{c}{Assumes mass ratio shown in column 6 and orbital period of 2000
d.}
\end{table}

\end{document}